\documentclass[prd,superscriptaddress,showpacs,showkeys]{revtex4-1}

\usepackage{listings}
\usepackage{mathtools}
\usepackage{amsfonts}
\usepackage{amsthm}
\usepackage{amssymb}
\usepackage{palatino}
\usepackage{eurosym}
\usepackage{amsmath}
\usepackage{epsfig}
\usepackage{graphics}
\usepackage{color}
\usepackage{graphicx}
\usepackage[font={footnotesize,it}]{caption}
\usepackage[colorlinks=true,
            linkcolor=blue,
            urlcolor=blue,
            citecolor=blue]{hyperref}
\makeindex

\begin{document}
\title{Hawking radiation and propagation of massive charged scalar field on a three-dimensional G\"{o}del black hole}

\author{P. A. Gonz\'{a}lez}
\email{pablo.gonzalez@udp.cl} \affiliation{Facultad de
Ingenier\'{i}a y Ciencias, Universidad Diego Portales, Avenida Ej\'{e}rcito
Libertador 441, Casilla 298-V, Santiago, Chile.}

\author{Ali \"{O}vg\"{u}n}
\email{ali.ovgun@pucv.cl}
\affiliation{Instituto de
F\'{i}sica, Pontificia Universidad Cat\'olica de Valpara\'{i}so,
Casilla 4950, Valpara\'{i}so, Chile.}

\affiliation{Physics Department, Arts and Sciences Faculty, Eastern Mediterranean University, Famagusta, North Cyprus via Mersin 10, Turkey}

\affiliation{Physics Department, California State University Fresno, Fresno, CA 93740,
USA.}
\affiliation{Stanford Institute for Theoretical Physics, Stanford University, Stanford,
CA 94305-4060, USA}

\author{Joel Saavedra}
\email{joel.saavedra@ucv.cl} \affiliation{Instituto de
F\'{i}sica, Pontificia Universidad Cat\'olica de Valpara\'{i}so,
Casilla 4950, Valpara\'{i}so, Chile.}

\author{Yerko V\'asquez}
\email{yvasquez@userena.cl}
\affiliation{Departamento de F\'isica y Astronom\'ia, Facultad de Ciencias, Universidad de La Serena,\\
Avenida Cisternas 1200, La Serena, Chile.}

\date{\today }

\begin{abstract}

In this paper we consider the three-dimensional G\"{o}del black hole as a background and we study the vector particle tunneling from this background in order to obtain the Hawking temperature. Then, we study the propagation of a massive charged scalar field and we find the quasinormal modes analytically, 
which turns out be unstable as a consequence of the existence of closed time-like curves.
Also, we consider the flux at the horizon and at infinity, and we compute the reflection and transmission coefficients as well as the absorption cross section. Mainly, we show that massive charged scalar waves can be superradiantly amplified by the three-dimensional G\"{o}del black hole and that the coefficients have an oscillatory behavior. Moreover, the absorption cross section is null at the high frequency limit and for certain values of the frequency. 

\end{abstract}

\maketitle

\tableofcontents
\newpage

\section{Introduction}
In 1974, Stephen Hawking showed that black holes evaporate and shrink because they emit radiation, today known as `Bekenstein-Hawking or Hawking radiation' \cite{Hawking:1974rv,Hawking:1974sw,Hawking:2016msc} with the contribution of Bekenstein \cite{Bekenstein:1974ax}. Nowadays, Hawking radiation is an important quantum effect of black hole physics, and it has been considered a principal tool in understanding the quantum nature of gravity. Moreover, it is determined by universal properties of the event horizon.  Apart from Hawking's original derivation, there are several approaches to obtain Hawking radiation, 
such as the approach developed by Christensen and Fulling \cite{Christensen:1977jc}, the quantum tunneling method, the null-geodesic method, the Hamilton-Jacobi method \cite{Banerjee:2008cf,Kerner:2006vu,Kerner:2007jk,Parikh:1999mf,Akhmedov:2006pg}  and the anomaly method developed by Robinson and Wilczek \cite{Robinson:2005pd}. These methods have acquired a growing interest and have been applied to several geometries that describe black holes. In addition, the tunneling method has recently been  used for the different type of spin/spinless particle such as a photon, vector particle, scalar particle, fermion, graviton or gravitino from a black hole or wormhole \cite{Akhmedova:2010zz,Kruglov1,Kruglov2,Kuang:2017sqa,Sakalli:2017ewb,Jusufi:2017trn,Sakalli:2016mnk,Ovgun:2016roz,Sakalli:2016cbo,Ovgun:2015box,Sakalli:2015jaa,Sakalli:2015taa,Akhmedova1,aji,Sakalli:2016jkf,Sakalli:2014wja,Sakalli:2016fif}. The modification of the Klein-Gordon and Dirac equations under the quantum gravity effect and their effect on the Hawking radiation from a black hole or wormhole by tunneling have also been investigated in many papers \cite{  Sakalli:2016mnk,Ovgun:2016roz,Ovgun:2015box}. On the other hand, Hawking radiation causes an unsolved paradox known as information loss \cite{Hawking:2016msc}. Maldacena  and Strominger showed that the Hawking radiation near the event horizon might be modified for a far observer from the black hole due to the greybody factors which modify the spectrum of emitted particles \cite{Maldacena:1996ix}, giving semiclassical features of the black holes which allow us a better understanding of the quantum nature of the black holes \cite{Harmark:2007jy}.
Moreover, it was shown that the Hawking radiation is connected with the quasinormal modes (QNMs) of black holes \cite{Corda:2012tz, Corda:2012dw}.

The QNMs were firstly studied a long time ago \cite{Regge:1957td, Zerilli:1971wd, Zerilli:1970se,
Kokkotas:1999bd, Nollert:1999ji, Konoplya:2011qq}. Using gravitational perturbations of a black hole spacetime, one can study the stability of the black hole. Also, it is possible to study the stability of the propagation of probe matter fields in the background of a black hole through the QNMs, which
have been obtained 
in several 
black hole geometries and also have acquired an important role in the
AdS/CFT correspondence \cite{Maldacena:1997re,Horowitz:1999jd}. Moreover, QNMs 
gives information about the spectrum of quantum area of the black hole horizon, mass and also entropy. Recently, since the detection of the gravitational waves from the merge of black holes, QNMs have gain more interest \cite{Abbott:2016blz}. This observation once again has proved the Einstein's gravity \cite{TheLIGOScientific:2016src}, but it leaves possibilities to other 
modified gravity theories
because there is large uncertainties in mass and angular momenta of the ringing black hole \cite{Konoplya:2016pmh}.

Lately, lower dimensional modified gravity models have gained many interest. One of them is the topologically massive gravity (TMG). In the model of TMG, the general theory of relativity is modified by adding the
Chern-Simons term to the action \cite{Deser:1981wh}. The main feature of the TMG is to give mass to the graviton. Moreover, a chiral theory of gravity at a special point can be constructed 
\cite{Deser:1982vy,Garbarz:2008qn,Nakasone:2009bn, Bergshoeff:2009aq, Oda:2009ys, Ohta:2011rv, Muneyuki:2012ur, Vasquez:2009mk}. In this paper, we 
consider a three-dimensional G\"{o}del black hole (GBH) \cite{Banados:2005da} in order to obtain the Hawking temperature from the vector particle tunneling and to study the propagation of a massive charged scalar field in this background.  As we will see, such propagation results unstable as a consequence of the existence of closed time-like curves (CTC). The effect  of CTC for string theory has been investigated in \cite{Biswas:2003ku,Brecher:2003rv,Brace:2003st,Takayanagi:2003ps}. Exact solutions for the QNMs of black holes in $2+1$ -dimensional spacetimes can be found in \cite{Cardoso:2001hn, Birmingham:2001pj, Konoplya:2004ik, Kwon:2011ey, CuadrosMelgar:2011up, Becar:2013qba, Gonzalez:2014voa, Catalan:2014una, Gonzalez:2017ptj}. Also, we find the 
greybody factor and we study the superradiance effect which is present in this background. The four-dimensional G\"{o}del spacetime is a exact solution of Einstein's gravity with
a large number of isometries and
was discovered by G\"{o}del in 1949 \cite{Godel:1949ga}.
Moreover the G\"{o}del spacetime  has a closed time-like curve through every point. The G\"{o}del black holes have been found \cite{Moussa:2008sj} in three spacetime dimensions, which are solutions to the Einstein-Maxwell-Chern Simons theory with a negative cosmological constant \cite{Banados:2005da}. The GBH is supported by the Abelian gauge field which also has a Chern-Simons interaction that produces the stress-tensor of a presureless perfect fluid, in analogy with five-dimensional G\"odel spacetimes and GBH found in \cite{Gauntlett:2002nw, Herdeiro:2002ft, Gimon:2003ms, Brecher:2003wq, Herdeiro:2003un, Behrndt:2004pn} in the context of 
supergravity theory. Interestingly, it is possible to relate, by means of T-duality, the G\"odel universes of \cite{Gauntlett:2002nw, Herdeiro:2002ft} to pp-waves \cite{Boyda:2002ba, Harmark:2003ud}. In \cite{Gimon:2003ms} a five-dimensional Schwarzschild black hole immersed in the G\"odel universe was found, which is stable against scalar field perturbations \cite{Konoplya:2005sy, Konoplya:2011ig}. On the other hand the three-dimensional GBH display
the same peculiar properties as their higher dimensional
counterparts \cite{Banados:2005da}. However, it is was shown that is not possible to obtain the absorption probability due to one cannot to find the conjugate charges associated with the left and the right temperatures in the CFT side \cite{Li:2012ee}. The conserved charges and thermodynamics of the GBH solution of five-dimensional minimal supergravity was studied in \cite{Barnich:2005kq}. Interestingly if the cosmological constant is too large in gauged supergravity, all closed time-like curve disappear. More recently, the QNMs and stability of a
five-dimensional rotating GBH were investigated
by Konoplya and Zhidenko \cite{Konoplya:2011it}, and other studies of stability in G\"odel-like solutions have been performed in \cite{Konoplya:2011ag}. The QNMs of the neutral scalar field of three-dimensional GBH were obtained by Li \cite{Li:2012qc}. In addition to the aforementioned solutions, GBH solutions in three dimensions in the presence of torsion was considered in \cite{Vasquez:2009mk}.

The paper is organized as follows. In Sec.~\ref{GBH} we give a brief review of a three-dimensional G\"{o}del black hole. In Sec.~\ref{Tunneling} we study the vector particle tunnelling from G\"{o}del spacetime and we obtain the Hawking temperature. In Sec.~\ref{QNMs} we solve analytically the Klein-Gordon equation for a massive charged scalar field and we find the QNMs. Then, in Sec.~\ref{Grey} we compute and analyze the reflection and transmission coefficients as well as the absorption cross section, and we find the condition for scalar waves to be superradiantly amplified by the black hole. We conclude with final remarks in Sec.~\ref{Final}.

\section{Three-dimensional G\"{o}del black hole}
\label{GBH}
The three-dimensional GBH is the solution of the Einstein-Maxwell-Chern-Simons theory described by the action:
 \begin{eqnarray}
 \label{action}
 S&=&\frac{1}{16\pi G}\int d^3 x\left[
 \sqrt{-g}\left( R+\frac{2}{\ell^2}
 -\frac{1}{4}F_{\mu\nu}F^{\mu\nu}\right)
 \right. \left.
 -\frac{\alpha}{2}\epsilon^{\mu\nu\rho}
 A_{\mu}F_{\nu\rho}\right]\;,
 \end{eqnarray}
where the cosmological constant is $\Lambda=-\frac{1}{\ell^2}$, with $\ell$ the AdS radius, and $G$ is the gravitational constant. 
The solutions for the metric and gauge potential of the action (\ref{action}) are given by \cite{Banados:2005da}:
 \begin{eqnarray}
 ds^2&=&(dt-2\alpha r d\varphi)^2
 -\Delta(r) d\varphi^2 +\frac{dr^2}{\Delta(r)}~,
 \end{eqnarray}
 \begin{equation}
 A_{\psi}=-\frac{4GQ}{\alpha}+\sqrt{1-\alpha^2l^2}\frac{2r}{\ell}~,
 \end{equation}
where the metric function is 
 \begin{eqnarray}
  \Delta(r)=(1+\alpha^2 \ell^2)\frac{2r^2}{\ell^2}
 -8G\nu r+\frac{4GJ}{\alpha}\;,
 \end{eqnarray}
and $J$ and $\nu$ are integration constants which depend on the angular momentum and mass of the GBH, while $Q$ is an arbitrary constant.
Note that the inner ($r_-$) and outer ($r_+$) event horizons of the GBH are calculated as follows:
 
 \begin{eqnarray}
 r_{\pm}=\frac{l^2}{1+\alpha^2 \ell^2}
 \left[ 2G\nu\pm\sqrt{4G^2\nu^2-
 \frac{2GJ}{\alpha}\frac{(1+\alpha^2 \ell^2)}{\ell^2}}
  \right]\;.
 \end{eqnarray}
 The thermodynamic properties of the GBH such as the Hawking temperature, the Bekenstein-Hawking entropy and the angular momentum are obtained  respectively as follows:
 \begin{eqnarray}
 \label{hr1}
 T_H&=&\frac{(1+\alpha^2 \ell^2)}{4\pi\alpha \ell^2}
 \frac{(r_+-r_-)}{r_+}\;,\nonumber\\
  S_{BH}&=&\frac{\pi\alpha r_+}{G}\;,\\\Omega_H&=&\frac{1}{2\alpha r_+}\;.\nonumber 
 \end{eqnarray}
 
 For simplicity, the metric of the GBH can be introduced in the dragging coordinate transformation as follows: $d\varphi=\frac{1}{2\alpha r}dt=\Omega dt$. Thus, we can avoid the dragging effect.
 The new form of the GBH metric without the dragging effect is written as:
 
  \begin{equation}
 ds^2=-F(r)dt^2+G(r)dr^2~, \label{proca2}
 \end{equation}
 where $F(r)=\frac{\Delta(r)}{4 \alpha^2 r^2} $ and $G(r)=\frac{1}{\Delta(r)}$.

\section{Vector particle tunneling from a three-dimensional G\"{o}del black hole}
\label{Tunneling}
In this section, we study the vector particle tunneling from the GBH in 2+1-dimensions to obtain the Hawking temperature. For this purpose, we use the Proca equation, which describes the spin-1 vector particles and we shall solve it using the semiclassical WKB approximation with the Hamilton-Jacobi method. The equation of the motion for the Proca field is given by \cite{Kruglov2} \begin{eqnarray}  \label{proca}
\frac{1}{\sqrt{-g}}\partial_{\mu}\left(\sqrt{-g}\,\Psi^{\nu\mu}\right)+\frac{%
m^{2}c^2}{\hbar^{2}}\Psi^{\nu}=0~,
\end{eqnarray}
where 
\begin{equation}
\Psi_{\mu\nu}=\partial_{\mu}\Psi_{\nu}-\partial_{\nu}\Psi_{\mu}~.  \label{13}
\end{equation}
Now, we solve the Proca equation on the background of the GBH given in Eq. (\ref{proca2}) and obtain the following equations:
\begin{eqnarray}
-\frac{1}{{h}^{2}\left[F\left(r\right)\right]^{2}\left[G\left(r\right)\right]^{2}}[{m}^{2}\psi_{1}\left(t,r\right)F\left(r\right)\left(G\left(r\right)\right)^{2}-G\left(r\right)F\left(r\right)\left({\frac{\partial^{2}}{\partial{r}^{2}}}\psi_{0}\left(t,r\right)\right){h}^{2} \notag \\
+G\left(r\right)F\left(r\right)\left({\frac{\partial^{2}}{\partial t\partial r}}\psi_{1}\left(t,r\right)\right){h}^{2}-G\left(r\right)\left({\frac{{\rm d}}{{\rm d}r}}F\left(r\right)\right)\left({\frac{\partial}{\partial t}}\psi_{1}\left(t,r\right)\right){h}^{2} \notag \\
+G\left(r\right)\left({\frac{{\rm d}}{{\rm d}r}}F\left(r\right)\right)\left({\frac{\partial}{\partial r}}\psi_{0}\left(t,r\right)\right){h}^{2}-F\left(r\right)\left({\frac{{\rm d}}{{\rm d}r}}G\left(r\right)\right)\left({\frac{\partial}{\partial t}}\psi_{1}\left(t,r\right)\right){h}^{2} \notag
\\ F\left(r\right)\left({\frac{{\rm d}}{{\rm d}r}}G\left(r\right)\right)\left({\frac{\partial}{\partial r}}\psi_{0}\left(t,r\right)\right){h}^{2}]=0 \label{eq1} \end{eqnarray} 
and
\begin{eqnarray}
{\frac {{m}^{2}\psi_{1} \left( t,r \right) F \left( r \right) -{h}^{2}{
\frac {\partial ^{2}}{\partial t\partial r}}\psi_{0} \left( t,r \right) +
{h}^{2}{\frac {\partial ^{2}}{\partial {t}^{2}}}\psi_{1} \left( t,r
 \right) }{{h}^{2}G \left( r \right) F \left( r \right) }}=0~. \label{eq2}
\end{eqnarray}
Then we apply the WKB approximation:
\begin{equation}
\Psi_{\nu}=C_{\nu}(t,r) e ^{\frac{i}{\hbar}\left(S_{0}(t,r)+\hbar\,S_{1}(t,r)+\hdots.\right)}~.
\end{equation}

Now, by using the Hamilton-Jacobi ansatz:
\begin{equation}
S_{0}(t,r)=-\epsilon t+W(r)+k,
\end{equation}
in which $\epsilon$ is the energy of the particle, in Eq. (\ref{eq1}) and Eq. (\ref{eq2}) and keeping only the leading order of $\hbar$, we obtain the following set of equations:
\begin{eqnarray}
{\frac { \left( -F \left( r \right)  \left( G \left( r \right) 
 \right) ^{2}{m}^{2}-F \left( r \right) G \left( r \right)  \left( {
\frac {\rm d}{{\rm d}r}}W \left( r \right)  \right) ^{2} \right) {\it 
C_{0}}}{ \left( F \left( r \right)  \right) ^{2} \left( G \left( r
 \right)  \right) ^{2}}}-{\frac {{\it C_{1}}\, \left( {\frac {\rm d}{
{\rm d}r}}W \left( r \right)  \right) \epsilon}{G \left( r \right) F
 \left( r \right) }}=0~, \end{eqnarray}
 
 \begin{eqnarray}
 -{\frac {{\it C_0}\, \epsilon{\frac {\rm d}{{\rm d}r}}W \left( r \right) }{G
 \left( r \right) F \left( r \right) }}+{\frac { \left( F \left( r
 \right) {m}^{2}-{ \epsilon}^{2} \right) {\it C_1}}{G \left( r \right) F
 \left( r \right) }=0}~.
\end{eqnarray}
 
The above equations can be written in a matrix form by considering $ \aleph(C_{0},C_{1})^{T}=0,$ where the matrix $\aleph$ is given by 

\begin{equation}
 \aleph=\left[ \begin {array}{cc} {\frac {-F \left( r \right)  \left( G
 \left( r \right)  \right) ^{2}{m}^{2}-F \left( r \right) G \left( r
 \right)  \left( {\frac {\rm d}{{\rm d}r}}W \left( r \right)  \right) 
^{2}}{ \left( F \left( r \right)  \right) ^{2} \left( G \left( r
 \right)  \right) ^{2}}}&-{\frac { \left( {\frac {\rm d}{{\rm d}r}}W
 \left( r \right)  \right) \epsilon}{G \left( r \right) F \left( r
 \right) }}\\ \noalign{\medskip}-{\frac { \left( {\frac {\rm d}{
{\rm d}r}}W \left( r \right)  \right) \epsilon}{G \left( r \right) F
 \left( r \right) }}&{\frac {F \left( r \right) {m}^{2}-{\epsilon}^{2}
}{G \left( r \right) F \left( r \right) }}\end {array} \right]~.
\end{equation}
Therefore, the non-trivial solution can be obtain from 
$\det\aleph=0$, which yields
\begin{equation}
-{\frac {{m}^{2} \left( F \left( r \right) G \left( r \right) {m}^{2}+
F \left( r \right)  \left( {\frac {\rm d}{{\rm d}r}}W \left( r
 \right)  \right) ^{2}-G \left( r \right) {\epsilon}^{2} \right) }{
 \left( G \left( r \right)  \right) ^{2} \left( F \left( r \right) 
 \right) ^{2}}}=0~,
\end{equation}
whose solution for the radial part is 
\begin{equation}
\label{int}
W(r)_{\pm}=\pm\int{{\frac{\sqrt{G(r)}}{\sqrt{F(r)}}\frac{\sqrt{\left(\epsilon^{2}-F(r){m}^{2}\right)}}{1}}}\mathrm{d}r~,
\end{equation}
where $F(r)=\frac{\lambda(r)}{4\alpha^{2}r^{2}}$ and $G(r)=\frac{1}{\lambda(r)}$. The positive/negative sign show the outgoing/ingoing spin-1 particles. Note that $F(r)\rightarrow0$ when $r \rightarrow r_h$.
To solve the integral (\ref{int}), first, the
function $F(r)$ is expanded in Taylor's series near the horizon 
\begin{equation}
F(r)\approx F(r_{h})+F^{\prime}(r_{h})(r-r_{h})+\frac{1}{2}F^{\prime\prime}(r_{h})(r-r_{h})^{2}~.\label{18}
\end{equation}
Then, the integral (\ref{int}) is evaluated around the pole, where there is the event horizon $r_{h}$, using the complex integral method and the result is found as follows:
\begin{equation}
ImW_{\pm}=\pm\frac{\pi \epsilon \alpha \ell^2 r_{+}}{(1+\alpha^2 \ell^2) (r_{+}-r_{-})}~. \label{19}
\end{equation}
Now, we set the probability of incoming spin-1 particles as $100\%$, where $P_{-}\simeq e^{-2ImW_{-}}=1$ to solve the factor two problem. Then, it leads to $Im \mathcal{S}_{-}=Im W_{-}+Im k=0$ for incoming, and  for outgoing it becomes $Im \mathcal{S}_{+}=Im W_{+}+Im k$. Hence, it is easy to see the relation that 
 $W_{+}=-W_{-}$.  It is worth noting that there is also a different way to solve it, which is described in Refs. \cite{Akhmedova:2010zz,Akhmedov:2006pg,Akhmedov:2006un}. Then, we obtain the correct tunneling probability of the outgoing spin-1 particles: 
\begin{equation}
P_{+}=e^{-2Im \mathcal{S}_{+}}\simeq e^{-4Im W_{+}}~,\label{23}
\end{equation}
and the corresponding tunneling rate is calculated as follows:
\begin{equation}
\Gamma=\frac{P_{+}}{P_{-}}\simeq e^{(-4Im W_{+})}~.\label{24}
\end{equation}
Afterwards, we compare the result with the Boltzmann formula $\Gamma=e^{-\beta E}$,
where $\beta$ is the inverse temperature, to find the Hawking temperature as
\begin{equation}
T_{H}=\frac{(1+\alpha^2 \ell^2)}{4\pi\alpha \ell^2}
 \frac{(r_+-r_-)}{r_+}~.\label{HRR}
\end{equation}
Therefore, the Hawking temperature obtained via the  tunneling method coincides with the Hawking temperature obtained from the surface gravity Eq. (\ref{hr1}).

\section{Quasinormal modes}
\label{QNMs}
In this section we calculate the QNMs of the three-dimensional GBH for a test charged massive scalar field. The Klein-Gordon equation in curved spacetime is given by 
\begin{equation}
\frac{1}{\sqrt{-g}}\left(\partial _{\mu }-iqA_{\mu}\right)\left( \sqrt{-g}g^{\mu \nu }\left(\partial
_{\nu }-iqA_{\nu}\right) \psi \right)=m^{2}\psi ~,  \label{KG}
\end{equation}%
where $m$ is the mass of the scalar field $\psi $. By means of the following ansatz 
\begin{equation} \label{phi}
\psi =e^{-i\omega t+ik \varphi}R(r)~,
\end{equation}%
the Klein-Gordon equation reduces to
\begin{equation}
\Delta \frac{d}{dr} \left( \Delta \frac{dR}{dr}\right)-\Delta \left( \omega^2+m^2\right)R+\left( 2 \alpha \omega r+q A_{\varphi}(r)-k\right)^2R  =0~.  \label{first}
\end{equation}%

\begin{figure}
\centering
\includegraphics[width=4.0in,angle=0,clip=true]{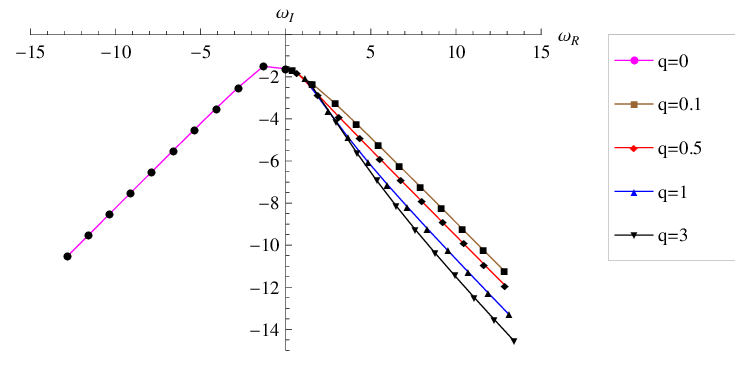}
\includegraphics[width=4.0in,angle=0,clip=true]{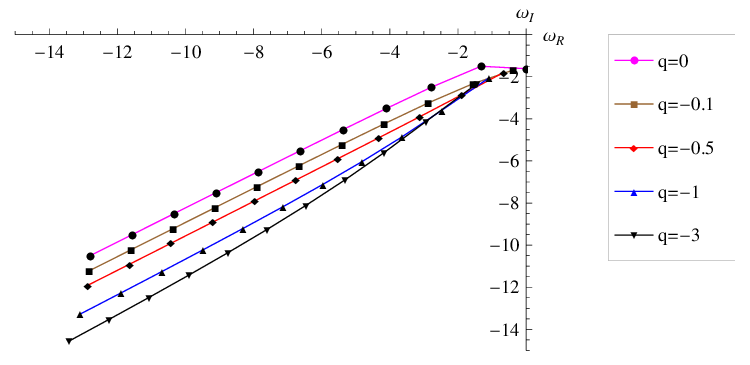}\caption{The behavior of QNFs, for different values of the charge of the scalar field $q$.; $\alpha=0.5$, $\nu=2$, $G=1$, $Q=1$, $J=1$, $\ell=1$, $k=0$, $m=1$ and $q=0,0.1,0.5,1,3$ (Top figure), $q=0,-0.1,-0.5,-1,-3$ (Bottom figure).}
\label{QNM22}
\end{figure}

\begin{figure}
\centering
\includegraphics[width=3.0in,angle=0,clip=true]{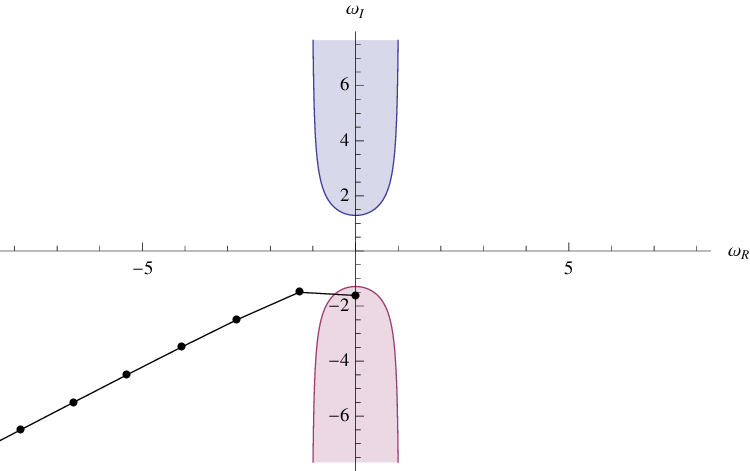}
\caption{First discrete set of QNFs and continuum QNFs (shaded region) for some values of the parameters: $\alpha=0.5$, $\nu=2$, $G=1$, $Q=1$, $J=1$, $\ell=1$, $m=1$ and $q=0$.}
\label{fig}
\end{figure}

\begin{figure}
\centering
\includegraphics[width=2.5in,angle=0,clip=true]{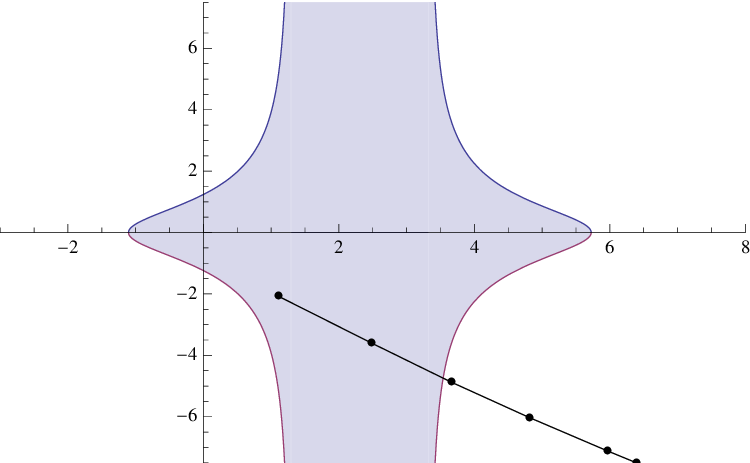}
\includegraphics[width=2.5in,angle=0,clip=true]{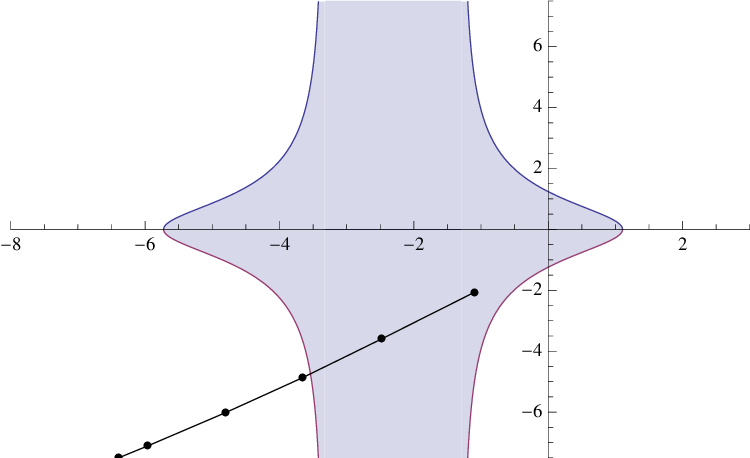}
\caption{First discrete set of QNFs and continuum QNFs (shaded region) for some values of the parameters: $\alpha=0.5$, $\nu=2$, $G=1$, $Q=1$, $J=1$, $\ell=1$, $m=1$, $q=2$ (left panel) and $q=-2$ (right panel).}
\label{fig1}
\end{figure}

\begin{figure}
\centering
\includegraphics[width=4.0in,angle=0,clip=true]{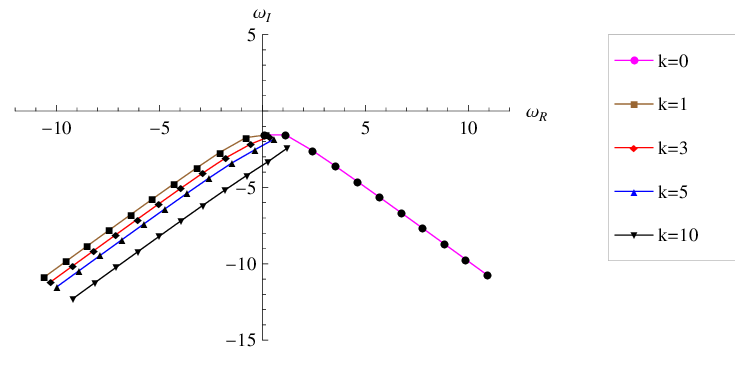}
\includegraphics[width=4.0in,angle=0,clip=true]{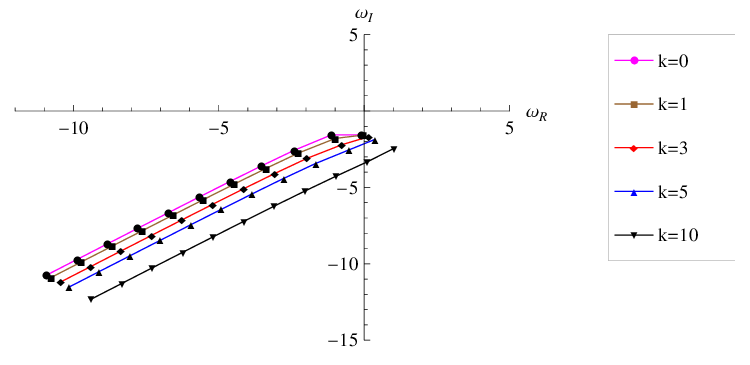}\caption{The behavior of QNFs, for different values of the angular momentum of the scalar field $k$.; $\alpha=0.5$, $\nu=2$, $G=1$, $Q=1$, $J=1$, $\ell=1$, $m=1$, $q=0.1$ (Top figure), $q=-0.1$ (Bottom figure) and $k=0,1,3,5,10$.}
\label{QNM1}
\end{figure}

\begin{figure}
\centering
\includegraphics[width=4.0in,angle=0,clip=true]{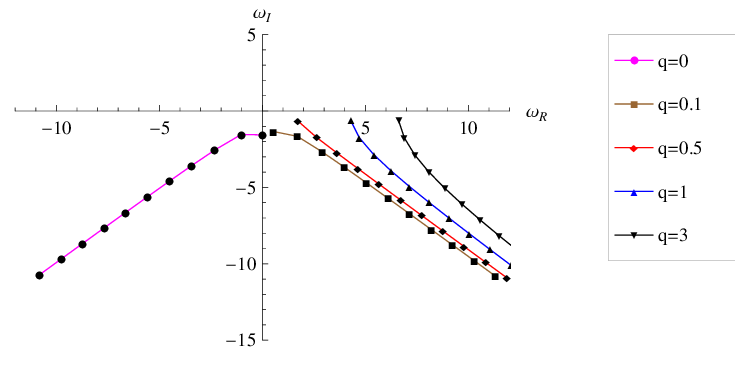}
º\includegraphics[width=4.0in,angle=0,clip=true]{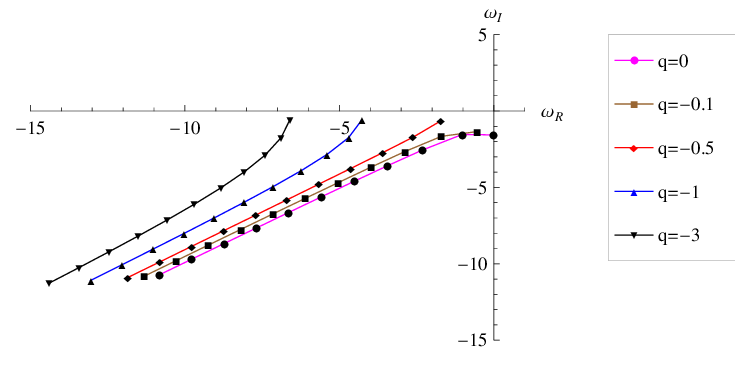}\caption{The behavior of QNFs, for different values of the charge of the scalar field $q$.; $\alpha=0.5$, $\nu=2$, $G=1$, $Q=1$, $J=1$, $\ell=1$, $k=0$, $m=1$ and $q=0,0.1,0.5,1,3$ (Top figure), $q=0,-0.1,-0.5,-1,-3$ (Bottom figure).}
\label{QNM2}
\end{figure}

Now, using the change of variable $z=\frac{r-r_+}{r-r_-}$, the Klein-Gordon equation (\ref{first}) can be written as
\begin{eqnarray}
\notag &&\lambda^2 (r_+-r_-)^2 z^2 R''(z)+\lambda^2(r_+-r_-)^2 z R'(z)+\Big((2 \alpha \omega+\frac{2q}{\ell}\sqrt{1-\alpha^2 \ell^2})^2 \left( \frac{r_+-zr_-}{1-z}\right)^2- \\
\notag &&\lambda z \frac{(r_+-r_-)^2}{(1-z)^2}(\omega^2+m^2)+(-4GQq/\alpha-k)(4 \alpha \omega+\frac{4q}{\ell}\sqrt{1-\alpha^2 \ell^2}) \left(\frac{r_+-zr_-}{1-z} \right)+ \\
&&(k+4GQq/\alpha)^2\Big)R =0~,
\end{eqnarray}
where $\lambda=\frac{2(1+\alpha^2 \ell^2)}{\ell^2}$ and if in addition we define $R(z)=z^{\tilde{\alpha}}(1-z)^{\tilde{\beta}}F(z)$, the above equation leads to the hypergeometric equation
\begin{equation}\label{hypergeometric}
 z(1-z)F''(z)+\left[c-(1+a+b)z\right]F'(z)-ab F(z)=0~,
\end{equation}
 where
\begin{equation}
\tilde{\alpha}_{\pm}=  \pm\frac{i\left(2 \alpha r_+ \omega + q A_{\varphi}(r_+) -k \right)}{\lambda (r_+-r_-)}~,
\end{equation}
\begin{equation}
\tilde{\beta}_{\pm}=\frac{1}{2}\pm \sqrt{\frac{1}{4}+\frac{\omega^2+m^2}{\lambda}-\frac{\left(2 \alpha \omega +\frac{2}{\ell} \sqrt{1-\alpha^2 \ell^2}q \right )^2}{\lambda ^2}}~,
\end{equation}
and the constants are given by
\begin{equation}\label{a}
a_{1,2}= \tilde{\alpha}+\tilde{\beta} \pm \frac{i\left(2 \alpha r_- \omega + q A_{\varphi}(r_-) -k \right)}{\lambda (r_+-r_-)}~,
\end{equation}
\begin{equation}
b_{1,2}= \tilde{\alpha}+\tilde{\beta} \mp \frac{i\left(2 \alpha r_- \omega + q A_{\varphi}(r_-) -k \right)}{\lambda (r_+-r_-)}~,
\end{equation}
\begin{equation}
c=1+2\tilde{\alpha}~.
\end{equation}
The general solution of the hypergeometric equation~(\ref{hypergeometric}) is
\begin{equation}
\label{HSolution}
F(z)=C_{1}{_2}F{_1}(a,b,c;z)+C_2z^{1-c}{_2}F{_1}(a-c+1,b-c+1,2-c;z)~,
\end{equation}
and it has three regular singular points at $z=0$, $z=1$, and
$z=\infty$. ${_2}F{_1}(a,b,c;z)$ is a hypergeometric function
and $C_{1}$ and $C_{2}$ are integration constants.
So, in the vicinity of the horizon, $z=0$ and using
the property $F(a,b,c,0)=1$, the function $R(z)$ behaves as
\begin{equation}\label{Rhorizon}
R(z)=C_1 e^{\tilde{\alpha} \ln z}+C_2 e^{-\tilde{\alpha} \ln z}~,
\end{equation}
and the scalar field $\varphi$, for $\tilde{\alpha}=\tilde{\alpha}_-$ can be written as follows:
\begin{equation}
\varphi\sim C_1 e^{-i\omega t -\frac{i\left(2 \alpha r_+ \omega + q A_{\varphi}(r_+) -k \right)}{\lambda (r_+-r_-)}\ln z}+C_2
e^{-i\omega t +\frac{i\left(2 \alpha r_+ \omega + q A_{\varphi}(r_+) -k \right)}{\lambda (r_+-r_-)}\ln z}~,
\end{equation}
in which the first term represents an ingoing wave and the second an outgoing wave in the black hole. So, by imposing that
only ingoing waves exist on the event horizon, this fixes  $C_2=0$. The radial
solution then becomes
\begin{equation}\label{horizonsolution}
R(z)=C_1 z^{\tilde{\alpha}} (1-z)^{\tilde{\beta} }{_2}F{_1}(a,b,c;z)=C_1z^{-\frac{i\left(2 \alpha r_+ \omega + q A_{\varphi}(r_+) -k \right)}{\lambda (r_+-r_-)}} (1-z)^{\tilde{\beta}}{_2}F{_1}(a,b,c;z)~.
\end{equation}
To implement boundary conditions at infinity ($z=1$), we
apply Kummer's formula
for the hypergeometric function \cite{Abramowitz}
\begin{equation}\label{relation}
{_2}F{_1}(a,b,c;z)=\frac{\Gamma(c)\Gamma(c-a-b)}{\Gamma(c-a)\Gamma(c-b)}F_1+(1-z)^{c-a-b}\frac{\Gamma(c)\Gamma(a+b-c)}{\Gamma(a)\Gamma(b)}F_2~,
\end{equation}
where
\begin{equation}
F_1={_2}F{_1}(a,b,a+b-c;1-z)~,
\end{equation}
\begin{equation}
F_2={_2}F{_1}(c-a,c-b,c-a-b+1;1-z)~.
\end{equation}
With this expression, the radial function~(\ref{horizonsolution}) reads
\begin{eqnarray}\label{R}\
R(z) &=& C_1 z^{-\frac{i\left(2 \alpha r_+ \omega + q A_{\varphi}(r_+) -k \right)}{\lambda (r_+-r_-)}}(1-z)^{\tilde{\beta}_+}\frac{\Gamma(c)\Gamma(c-a-b)}{\Gamma(c-a)\Gamma(c-b)} F_1+ \\
&& C_1 z^{-\frac{i\left(2 \alpha r_+ \omega + q A_{\varphi}(r_+) -k \right)}{\lambda (r_+-r_-)}}(1-z)^{\tilde{\beta}_-}\frac{\Gamma(c)\Gamma(a+b-c)}{\Gamma(a)\Gamma(b)}F_2~,
\end{eqnarray}
and at infinity it can be written as
\begin{equation}\label{R2}\
R_{asymp}(z) = C_1 (1-z)^{\tilde{\beta}_+}\frac{\Gamma(c)\Gamma(c-a-b)}{\Gamma(c-a)\Gamma(c-b)}+C_1(1-z)^{\tilde{\beta}_-}\frac{\Gamma(c)\Gamma(a+b-c)}{\Gamma(a)\Gamma(b)}~.
\end{equation}
We notice that $\tilde{\beta}_-$ can have a positive or negative real part. So, for negative values of the real part of $\tilde{\beta}_-$,
the scalar field at infinity vanishes
if $a=-n$ or $b=-n$ for $n=0,1,2,...$. 
Therefore, the first discrete set of QNFs corresponding to $a=-n$ is given by
\begin{equation}
\omega = -\sqrt{-\frac{2q}{\ell} i (1+2n) \sqrt{1-\alpha ^2 \ell^2}-\alpha ^2(1+2n)^2+ \lambda n (1+n)-m^2}- i\alpha  (1+2n)~,
\end{equation}
which coincide with the results found by  Li \cite{Li:2012qc} for a neutral scalar field. Also, it can be rewritten as
\begin{eqnarray}
\notag \omega &=& sgn(q)\frac{1}{\sqrt{2}}\Big[-\left( m^2 +\alpha^2 (1+2 n)^2 - \lambda n(1+n)\right) + \Big(\left( m^2 +\alpha^2 (1+2 n)^2 - \lambda n(1+n)\right)^2 \\
\notag && + 4 q^2(1- \alpha^2 \ell^2) (1+2n)^2 / \ell^2  \Big)^{1/2} \Big]^{1/2} - i \frac{1}{\sqrt{2}}\Big[ m^2 +\alpha^2 (1+2 n)^2 - \lambda n(1+n) + \\
\notag && \Big(\left( m^2 +\alpha^2 (1+2 n)^2 - \lambda n(1+n)\right)^2 + 4 q^2 (1- \alpha^2 \ell^2) (1+2n)^2  /\ell^2  \Big)^{1/2} \Big]^{1/2}-i \alpha (1+2n)~, \\
\end{eqnarray}
where we have separated the real and the imaginary part of the QNFs and $sgn()$ refers to the sign function. Note that the above expression does not depend on the field angular momentum $k$. In Fig. \ref{QNM22} we plot the behavior of some QNFs for positive and negative values of the charge of the scalar field and $n=0,1,...,10$. We observe that the discrete QNFs have a negative imaginary part and the real part is positive for positive values of the charge of the scalar field and negative for negative values of the charge. 
The second discrete set of QNFs is given by $b=-n$ and it is given by
\begin{eqnarray}
\omega &=& \frac{\alpha ^2 l \left(-16 \alpha  q r_- r_+ \sqrt{1-\alpha ^2 l^2}+l (r_-+r_+) (16 G q Q+4 \alpha  k+i \alpha  \lambda  (2 n+1) (r_--r_+))\right)}{\alpha ^2 l^2 \left(16 \alpha ^2 r_- r_++\lambda  (r_--r_+)^2\right)}\\
\notag &&-\frac{\sqrt{\alpha ^2 l^2 (r_--r_+)^2 \left(8 \alpha  G l q Q \left(\alpha ^2 \left(2 B k l+4 k^2 l^2 \left(4 \alpha ^2-\lambda \right)+\text{$\lambda $D}\right)+B+4 k l \left(4 \alpha ^2-\lambda \right)\right)+64 G^2 l^2 q^2 Q^2 \left(4 \alpha ^2-\lambda \right)\right)}}{\alpha ^2 l^2 \left(16 \alpha ^2 r_- r_++\lambda  (r_--r_+)^2\right)}~,
\end{eqnarray}
where
\begin{equation}
B=\lambda  \left(4 q \sqrt{1-\alpha ^2 l^2} (r_-+r_+)+i l (2 n+1) \left(4 \alpha ^2-\lambda \right) (r_--r_+)\right)~,
\end{equation}
\begin{eqnarray}
\notag C&=&-\lambda  \left(m^2 (r_--r_+)^2+\alpha ^2 \left(-2 (4 n (n+1)-1) r_- r_++(2 n r_-+r_-)^2+(2 n r_++r_+)^2\right)\right)\\
&&-16 \alpha ^2 m^2 r_- r_++\lambda ^2 n (n+1) (r_--r_+)^2~,
\end{eqnarray}
and
\begin{equation}
D=C l^2+2 i \lambda  l (2 n+1) q \sqrt{1-\alpha ^2 l^2} (r_--r_+) (r_-+r_+)+16 q^2 r_- r_+ \left(\alpha ^2 l^2-1\right)~.
\end{equation}
Now, we show some QNFs in the Figs. \ref{QNM1} and \ref{QNM2} for some values of the parameters.  In Fig. \ref{QNM1} we consider different values for the angular momentum of the scalar field and in Fig. \ref{QNM2} we consider different values for the charge of the scalar field. We observe that the discrete QNFs have a negative imaginary part and the real part is positive and negative for different values of the angular momentum of the scalar field. We also observe that the discrete QNFs have a negative imaginary part and the real part is positive for positive values of charge scalar field and negative for negative values of charge scalar field, which correspond to the same behavior previously described for the first set of QNFs. Note that for a large overtone number $n$ the QNFs present an imaginary part whose absolute value is greater than for a small overtone number (see Fig. \ref{QNM22}, \ref{QNM1} and \ref{QNM2}).  Additionally, we note that $\beta_+$ and $\beta_-$ can have both a positive real part for a continuum range of values of $\omega$; thus, the scalar field is null at spatial infinity for those values of $\omega$.  The imaginary part of these continuum frequencies can be positive; therefore, the propagation of a massive charged scalar field on a three-dimensional GBH is unstable.  
In fact, notice that $\tilde{\beta}_{\pm}$ can be written as
\begin{equation}
\tilde{\beta}_{\pm}=\frac{1}{2}\pm (A+iB)\, ,
\end{equation}
where
\begin{eqnarray}
A &=& \sqrt{\frac{z_1+\sqrt{z_1^2+z_2^2}}{2}} \, , \\ 
B &=& \frac{z_2}{\sqrt{2 (z_1+\sqrt{z_1^2+z_2^2})}}\, ,
\end{eqnarray}
and
\begin{eqnarray}
z_1 &=& \frac{1}{4}+(\omega_R^2-\omega_I^2)(\frac{1}{\lambda}-4\frac{\alpha^2}{\lambda^2})+\frac{m^2}{\lambda}-\frac{8 \alpha q}{\ell \lambda^2} \sqrt{1-\alpha^2 \ell^2} \omega_R-\frac{4 q^2}{\ell^2 \lambda^2}(1-\alpha^2 \ell^2) \, , \\ 
z_2 &=& 2 \omega_R \omega_I (\frac{1}{\lambda}-4\frac{\alpha^2}{\lambda^2})-\frac{8 \alpha q}{\ell \lambda^2} \sqrt{1-\alpha^2 \ell^2} \omega_I \, ,
\end{eqnarray}
where $\omega_R$ and $\omega_I$ denote the real and imaginary part of $\omega$ respectively.
The real part of $\tilde{\beta_+}$  is always positive; however, the real part of $\tilde{\beta}_-$ is positive for $\frac{1}{2}-A>0$, in this case there is a continuum of QNFs. In Figs. (\ref{fig}) and (\ref{fig1}) we show the first discrete set of QNFs and  continuum QNFs (shaded region) for some values of the parameters.


A relation between unstable QNMs and closed time-like curves was found in the rotating infinity cylinder space-times in general relativity, where was concluded that the infinite cylinders that have closed time-like curves are unstable against scalar perturbations \cite{Pavan:2009wt}; therefore, one can expect that there is a deep relationship between the existence of instability and the existence of closed time-like curves.

\begin{figure}[!h]
	\begin{center}
		\includegraphics[width=60mm]{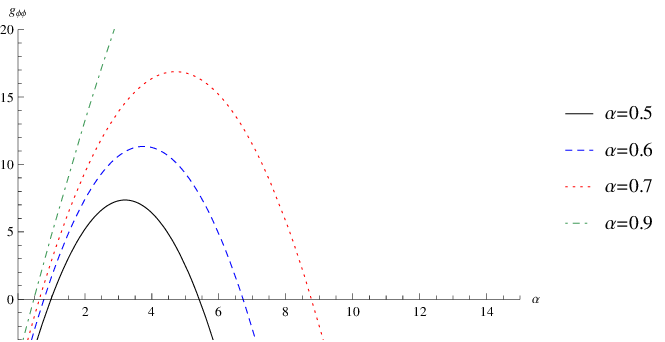}
        \includegraphics[width=60mm]{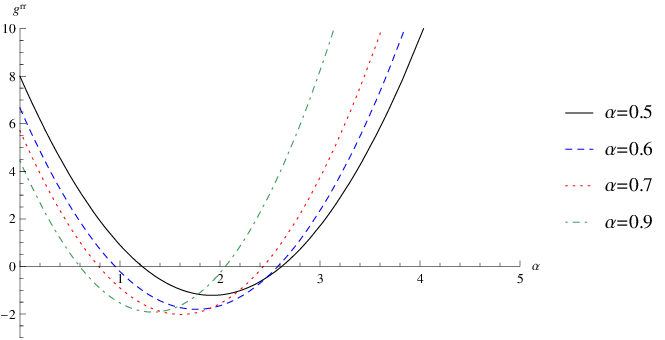}
	\end{center}
	\caption{Normal region ($g_{\phi \phi}(r)>0$) and metric function $g_{rr}$ for $G=J=\ell=1$, $\nu$=1.2 and different values of $\alpha$.}
	\label{f1}
\end{figure}

\begin{figure}[!h]
	\begin{center}
		\includegraphics[width=60mm]{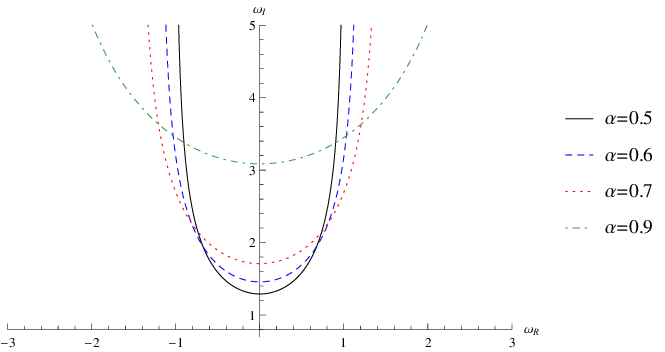}
	\end{center}
	\caption{region of instability (region above the curves) for $q$=0, $\ell=1$, $m$=1 and different values of $\alpha$.}
	\label{f2}
\end{figure}

In Fig. (\ref{f1}) we plot the behavior of the normal region and the metric function $g^{rr}$ for some values of the parameters and different values of the parameter $\alpha$, and in Fig. (\ref{f2}) we plot the behavior of the unstable QNMs (region above the curves) for the same values of the parameter $\alpha$ of Fig. (\ref{f1}). For simplicity we have set $q=0$ (in this case the region of instability just depends on the values of $m$, $\alpha$ and $\ell$). In Fig. (\ref{f1}) we observe that the size of the normal region increases with $\alpha$, and in Fig. (\ref{f2}) we observe that the lowest imaginary part of the unstable modes (the point where the curves intersect the $\omega_I$ axis) also increases with $\alpha$, and in the limit $\alpha \rightarrow 1$ (notice that for $\alpha^2 \ell^2=1$ the metric reduces to the BTZ black hole), the normal region extends to infinity and the instability region is pushed to infinity. Based in this results, we are tempted to think that there exist a straightforward correlation between the existence of unstable modes and the existence of closed time-like curves; however, it is worth to mention that in the five-dimensional G\"odel black hole of \cite{Gimon:2003ms} no evidence of instability was found in the quasinormal frequencies for a test scalar field \cite{Konoplya:2005sy, Konoplya:2011ig}.

\newpage

\section{Reflection and transmission coefficients and absorption cross section}
\label{Grey} 

From the Klein-Gordon equation in curved space is possible to obtain the conserved current $j^{\mu}$ for scalar particles, with $\partial_{\mu} j^{\mu} =0$, as
\begin{equation}
j^{\mu}=\frac{1}{2 i} \sqrt{-g} \left( \phi^{\ast} \partial^{\mu} \phi -\phi \partial^{\mu} \phi \right)\, .
\end{equation}
The flux can be obtained from the radial component of this current, and using the ansatz \eqref{phi} for the scalar field, yields 
\begin{equation} \label{fluxd}
\mathcal{F}=\frac{1}{2 i} \sqrt{-g} g^{rr} \left( R^{\ast} \partial_{r} R -R \partial_{r} R \right) .
\end{equation}
Using this flux, we can obtain the reflection and the transmission coefficients, which  are given by \cite{ch1,ch2}
\begin{equation}\label{reflectiond}\
\mathcal{R}=\left|\frac{\mathcal{F}_{\mbox{\tiny asymp}}^{\mbox{\tiny
out}}}{\mathcal{F}_{\mbox{\tiny asymp}}^{\mbox{\tiny in}}}\right|~, \qquad
\mbox{and} \qquad  \mathcal{T} = \left|\frac{\mathcal{F}_{\mbox{\tiny
hor}}^{\mbox{\tiny in}}}{\mathcal{F}_{\mbox{\tiny asymp}}^{\mbox{\tiny
in}}}\right|~,
\end{equation}
where $\mathcal{F}^{\mbox{\tiny in}}_{\mbox{\tiny asymp}}$ is the incident flux in the asymptotic region,  $\mathcal{F}^{\mbox{\tiny out}}_{\mbox{\tiny asymp}}$ is the reflected flux in the asymptotic region and $\mathcal{F}^{\mbox{\tiny in}}_{\mbox{\tiny hor}}$ is the transmitted flux to the black hole. So, in order to calculate the above coefficients we need to know the behavior
of the radial function both on the horizon and at asymptotic
infinity. The behavior at the horizon is given by Eq. (\ref{Rhorizon}) with $C_2=0$, choosing the negative value of $\tilde{\alpha}$ and using Eq. (\ref{fluxd}), we get the following flux on the horizon:
\begin{equation}
\mathcal{F}
_{\mbox{\tiny hor}}^{\mbox{\tiny in}}=-|C_1|^2 (2 \alpha r_+ \omega +qA_{\varphi}(r_+)-k)~.
\end{equation}
On the other hand, by applying Kummer's formula (\ref{relation})  for the hypergeometric function in Eq. (\ref{HSolution}), the asymptotic behavior of $R(z)$  can be written as
\begin{equation}
R\left( z \rightarrow 1 \right) = D_1 (z-1)^{\beta}+D_2 (z-1)^{1-\beta}~,
\end{equation}
where
\begin{eqnarray}
D_1& = & C_{1}\frac{\Gamma \left( c\right) \Gamma \left( c-a-b\right) }{\Gamma \left(
c-a\right) \Gamma \left( c-b\right) }~, \notag \\
D_2& = & C_{1}\frac{\Gamma \left( c\right) \Gamma \left( a+b-c\right) }{\Gamma \left( a\right) \Gamma \left( b\right) }~.
\end{eqnarray}
Thus, using Eq. (\ref{fluxd}) we obtain the flux at infinity
\begin{equation}
\mathcal{F}_{\mbox{\tiny asymp}}=2 \lambda (r_+-r_-)\sqrt{\frac{1}{4}+\frac{\omega^2+m^2}{\lambda}-\frac{\left(2 \alpha \omega +\frac{2}{\ell} \sqrt{1-\alpha^2 \ell^2}q \right )^2}{\lambda ^2}}(-|A_1|^2+|A_2|^2 )~,
\end{equation}
for $\beta=\beta_{+}$, where $A_1=\frac{1}{2}(D_1+iD_2)$ and $A_2=\frac{1}{2}(D_1-iD_2)$. Therefore, the reflection and transmission coefficients are given by \cite{Sakalli:2016fif}
\begin{equation}
\label{RE}
\mathcal{R}=\frac{|A_2|^2 }{|A_1|^2 }~,
\end{equation}
\begin{equation}
\label{TE}
\mathcal{T}=\frac{|C_1|^2 (2 \alpha r_+ \omega +qA_{\varphi}(r_+)-k)}{2 \lambda (r_+-r_-)\sqrt{\frac{1}{4}+\frac{\omega^2+m^2}{\lambda}-\frac{\left(2 \alpha \omega +\frac{2}{\ell} \sqrt{1-\alpha^2 \ell^2}q \right )^2}{\lambda ^2}}|A_1|^2}~,
\end{equation}
and the absorption cross section, $\sigma_{abs}$, is given by \cite{Gubser:1997qr}
\begin{equation}\label{absorptioncrosssection}\
\sigma_{abs}=\frac{\mathcal{T}}{\omega}=\frac{|C_1|^2 (2 \alpha r_+ \omega +qA_{\varphi}(r_+)-k)}{2 \lambda (r_+-r_-)\sqrt{\frac{1}{4}+\frac{\omega^2+m^2}{\lambda}-\frac{\left(2 \alpha \omega +\frac{2}{\ell} \sqrt{1-\alpha^2 \ell^2}q \right )^2}{\lambda ^2}}|A_1|^2 \omega}~.
\end{equation}

It should be mentioned that one way to find the conditions for superradiance amplification of a scatter wave is to compute the greybody factor and the reflection coefficients. Then, if the greybody factor is negative or the reflection coefficient is greater than 1, then the scalar waves can be superradiantly amplified by the black hole \cite{Benone:2015bst, Gonzalez:2017shu}. So, this condition implies that $2 \alpha \omega r_++q A_{\varphi}(r_+)-k<0$.

\begin{figure}
\centering
\includegraphics[width=4.0in,angle=0,clip=true]{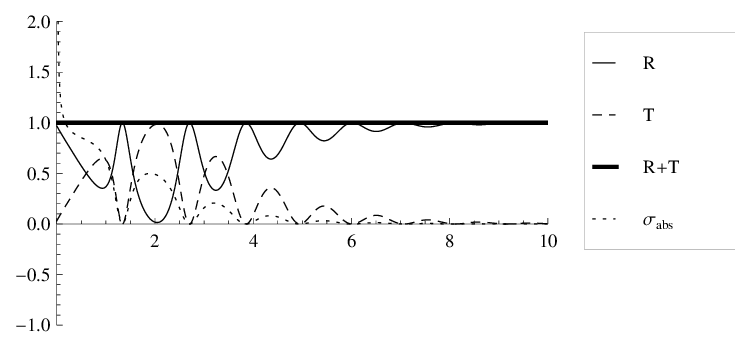}
\includegraphics[width=4.0in,angle=0,clip=true]{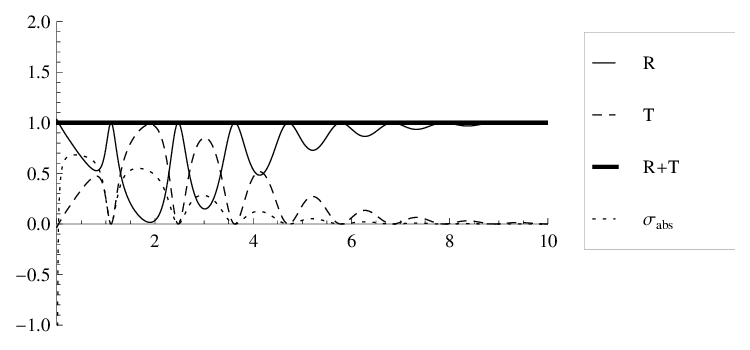}
\caption{The reflection coefficient $R$ (solid curve), the transmission coefficient $T$ (dashed curve), $R+T$ (thick curve) and the absorption cross section $\sigma_{abs}$ (dotted curve) as a function of $\omega$, for $\alpha=0.5$, $\nu=2$, $m=1$, $q=0.1$ (Top figure), $q=-0.1$ (Bottom figure) , $G=1$, $Q=1$, $J=1$, $k=0$ and $\ell=1$.}
\label{coef}
\end{figure}

\begin{figure}
\centering
\includegraphics[width=4.0in,angle=0,clip=true]{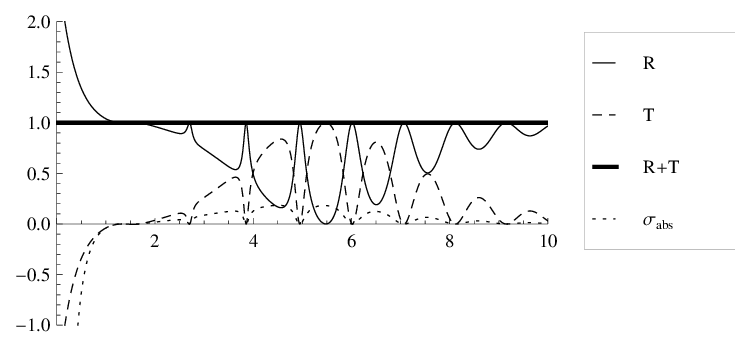}
\includegraphics[width=4.0in,angle=0,clip=true]{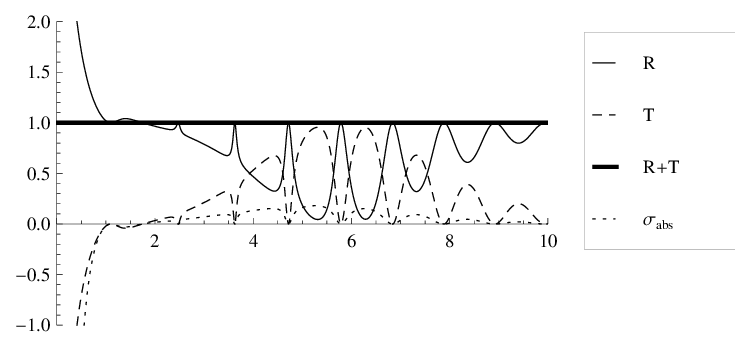}
\caption{The reflection coefficient $R$ (solid curve), the transmission coefficient $T$ (dashed curve), $R+T$ (thick curve) and the absorption cross section $\sigma_{abs}$ (dotted curve) as a function of $\omega$, for $\alpha=0.5$, $\nu=2$, $m=1$, $q=0.1$ (Top figure), $q=-0.1$ (Bottom figure), $G=1$, $Q=1$, $J=1$, $k=10$ and $\ell=1$.}
\label{coeff}
\end{figure}

Then we can numerically study the reflection coefficient (\ref{RE}), transmission coefficient~(\ref{TE}) and absorption cross section~(\ref{absorptioncrosssection}) of the three-dimensional GBH for charged massive scalar fields for different values of the parameters. Therefore, the reflection and transmission coefficients and the absorption cross section in Figs.~(\ref{coef}- \ref{coeff}) are plotted for massive charged scalar fields with $m=1$ and with a positive charge $(q=0.1)$ and a negative charge $(q=-0.1)$. Essentially, we observe in Fig. (\ref{coef}), where we have considered radial scalar field $(k=0)$,  that the reflection coefficient is 1 at the low frequency limit, then acquires an oscillatory behavior, reaching a minimum value for $\omega \approx 2.04$ when q=0.1 and $\omega \approx 1.90$ when $q=-0.1$, and for the high frequency limit this coefficient tends to $1$. The behavior of the transmission coefficient is opposite to the behavior of $\mathcal{R}$, with $\mathcal{R}+\mathcal{T} = 1$. In addition, the absorption cross section is not null and it diverges in the low-frequency limit and tends to zero in the high-frequency limit. It is worth to mention that for certain values of the frequency $\omega \approx 1.34, 2.70, 3.86, 4.95, 6.02, 7.07, 8.11$, for $q=0.1$, and $\omega \approx 0.04, 1.11, 2.47, 3.62, 4.72, 5.79, 6.84, 7.88, 8.92$, for $q=-0.1$, the absorption cross section is null; this oscillatory behavior has not been observed in other geometries, (see for instance \cite{Gonzalez:2010vv,Gonzalez:2010ht,Gonzalez:2011du,Catalan:2014ama,Becar:2014aka,Becar:2014saa}).  The discrete values of $\omega$ for which the transmission coefficient and the absorption cross section are zero can be found from the condition $c-a-b=-n$, which yields
\begin{equation}
\omega=\frac{8 q \ell \alpha+ \sqrt{\lambda \ell^2 (16 q^2+2 \lambda (n^2-1)-8 m^2)}}{4 \sqrt{1-\alpha^2 \ell^2}} \,,
\end{equation}
where $n$ is a positive integer number or zero that must satisfy the condition $n^2>1+4( m^2-2 q^2)/ \lambda$ to guarantee a real value of $\omega$.

Then, in Fig. (\ref{coeff}), we consider a scalar field with angular momentum $(k=10)$ and we observe that the reflection coefficient is greater than 1 at the low frequency limit for $\omega < 1.67$ when q=0.1 and $\omega < 1.75$ when $q=-0.1$, which corresponds to the superradiant regimes. Then, for frequencies $\omega > 1.67$ when q=0.1 and $\omega > 1.75$ when $q=-0.1$, the reflection coefficient becomes smaller than $1$ and acquires an oscillatory behavior, reaching a minimum value for $\omega \approx 5.48$ and null when q=0.1. For $q=-0.1$ the minimum value is not null; and for the high frequency limit it tends to $1$. The behavior of the transmission coefficient is opposite to the behavior of $\mathcal{R}$, with $\mathcal{R}+\mathcal{T} = 1$. In addition, the absorption cross section is negative in the superradiant regime and tends to zero in the high-frequency limit.

\begin{figure}
\centering
\includegraphics[width=4.0in,angle=0,clip=true]{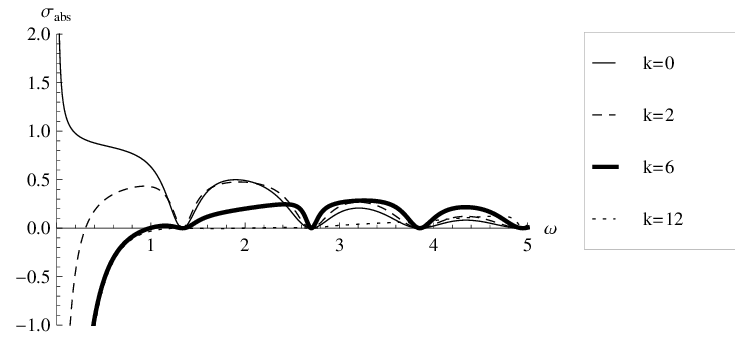} 
\includegraphics[width=4.0in,angle=0,clip=true]{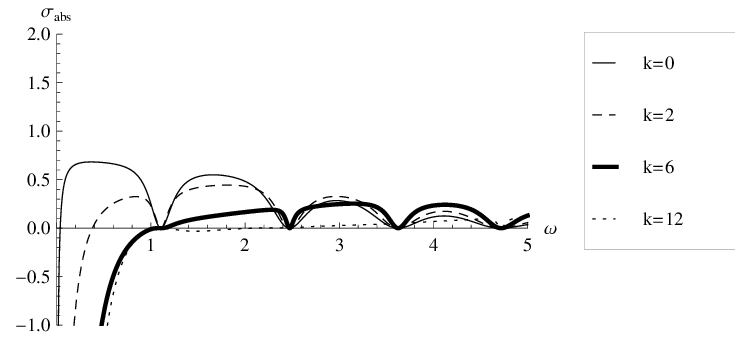}
\caption{The behavior of $\sigma_{abs}$ as a function of $\omega$, for different values of the angular momentum of the scalar field $k$; $\alpha=0.5$, $\nu=2$, $m=1$, $q=0.1$ (Top figure), $q=-0.1$ (Bottom figure), $G=1$, $Q=1$, $J=1$, $\ell=1$ and $k=0,2,6,12$.}
\label{Sigmak}
\end{figure}

\begin{figure}
\centering
\includegraphics[width=4.0in,angle=0,clip=true]{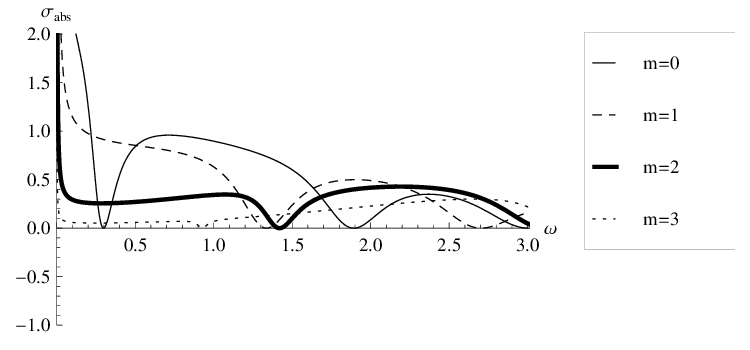}
\includegraphics[width=4.0in,angle=0,clip=true]{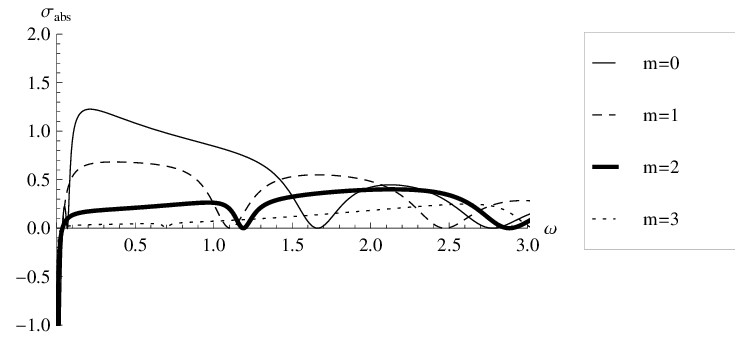}
\caption{The behavior of $\sigma_{abs}$ as a function of $\omega$, for different values of the mass of the scalar field $m$; $\alpha=0.5$, $\nu=2$, $q=0.1$ (Top figure), $q=-0.1$ (Bottom figure), $G=1$, $Q=1$, $J=1$, $\ell=1$, $k=0$, and $m=0,1,2,3$.}
\label{Sigmam}
\end{figure}

\begin{figure}
\centering
\includegraphics[width=4.0in,angle=0,clip=true]{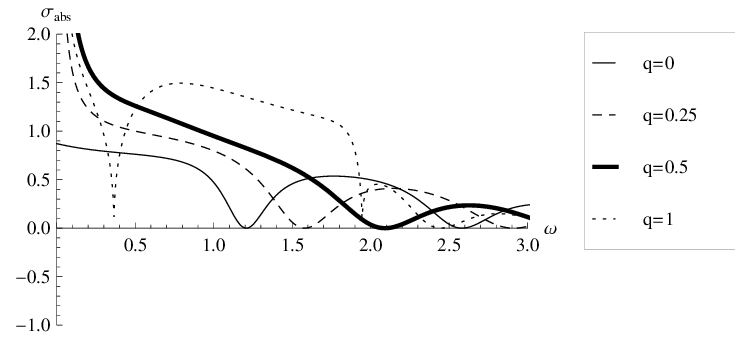}
\includegraphics[width=4.0in,angle=0,clip=true]{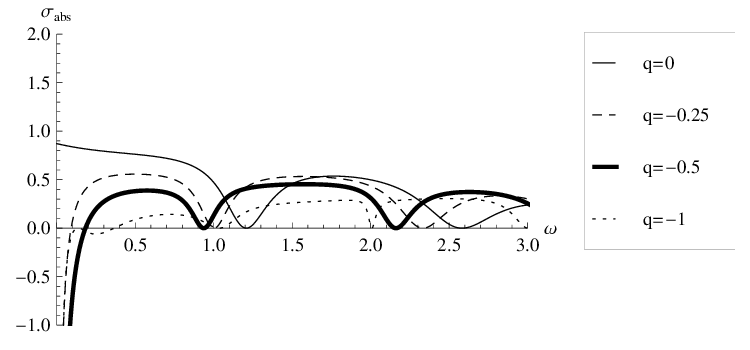}
\caption{The behavior of $\sigma_{abs}$ as a function of $\omega$, for different values of the charge of the scalar field $q$; $\alpha=0.5$, $\nu=2$, $m=1$, $G=1$, $Q=1$, $J=1$, $\ell=1$, $k=0$ and $q=0, 0.25,0.5,1$ (Top figure), $q=0, -0.25,-0.5,-1$ (Bottom figure).}
\label{Sigmaq}
\end{figure}

Finally, in Fig. (\ref{Sigmak}) we show the behavior of $\sigma_{abs}$ for different values of angular momentum of the scalar field $k$ and we observe that the values of the frequency $\omega$, for which the absorption cross section being null does not depend on the angular momentum of the scalar field. However, these frequency values depend on the mass of the scalar field, see Fig.~(\ref{Sigmam}), and on the charge of the scalar field, see Fig.~(\ref{Sigmaq}). Also, it is worth mentioning that for the radial and uncharged scalar field the absorption cross section is finite in the low-frequency limit and is given by $\sigma_{abs} \approx  0.87$, see Fig.~(\ref{Sigmaq}). 

\newpage

\section{Final remarks}
\label{Final}

In this manuscript we have studied the Hawking radiation for vector particles tunneling from the three-dimensional GBH and the propagation of charged scalar field perturbations in this background. We have obtained analytical expressions for the QNFs, by imposing Dirichlet boundary conditions, and also for the reflection $\mathcal{R}$ and transmission $\mathcal{T}$ coefficients and for the absorption cross section. First, we have found that the correct Hawking temperature is recovered. 
Then, we found that discrete sets of QNFs have a negative imaginary part. However, there is a range of continuous QNFs that have a positive imaginary part, implying that the propagation of a charged scalar field on a three-dimensional GBH is unstable as a consequence of the existence of closed time-like curves.\\

Also, we have shown that the superradiance effect is present in the propagation of massive charged scalar field in the three-dimensional GBH, when $2 \alpha \omega r_++q A_{\varphi}(r_+)-k<0$. Under this regime the greybody factor is negative or the reflection coefficient is greater than 1, then the charged massive scalar waves can be superradiantly amplified by the black hole. Outside this superrradiance regime the reflection coefficient starts at $1$ then becomes smaller than $1$ and acquires an oscillatory behavior, reaching a minimum value which is null for a positive charge of the scalar field and not null for a negative charge of the scalar field and for a high frequency limit tends to $1$. The behavior of the transmission coefficient is opposite to the behavior of the reflection coefficient, with $\mathcal{R}+\mathcal{T} = 1$.\\

Furthermore, as we pointed out, the absorption cross section is negative in the superradiant regime and tends to zero in the high-frequency limit. Also, the absorption cross section acquires an oscillatory behavior and  is null for certain values of the frequency depending on the mass and charge of the scalar field, but not depending on the angular momentum of the scalar field. Therefore, a wave emitted from the horizon with these values of frequencies or with high frequency does not reach the spatial infinity and is totally reflected, because the fraction of particles penetrating the potential barrier vanishes. It is worth noting that the absorption cross section is finite in the low-frequency limit for radial and uncharged scalar field.

\begin{acknowledgments}
This work was partially supported by the Comisi\'{o}n
Nacional de Ciencias y Tecnolog\'{i}a through FONDECYT Grants No. 3170035 (A\"{O}) and 11140674 (PAG) and by the Direcci\'{o}n de Investigaci\'{o}n y Desarrollo de la Universidad de La Serena (Y.V.). P. A. G. acknowledges the hospitality of the Universidad de La Serena, National Technical University of Athens and Pontificia Universidad Cat\'{o}lica de Valpara\'{i}so where part of this work was undertaken. A. \"{O}. is grateful to Prof. Douglas Singleton for hosting him as a
research visitor at the California State University, Fresno. In addition A. 
\"{O}. would like to thank Prof. Leonard Susskind and Stanford Institute for
Theoretical Physics for hospitality.
\end{acknowledgments}


\begin{thebibliography}{99}



\bibitem{Hawking:1974rv}  S.~W.~Hawking,  
Nature \textbf{248}, 30 (1974).

\bibitem{Hawking:1974sw}  S.~W.~Hawking,  
Commun.\ Math.\ Phys.\ \textbf{43}, 199 (1975)  Erratum: [Commun.\ Math.\
Phys.\ \textbf{46}, 206 (1976)].


\bibitem{Hawking:2016msc}  S.~W.~Hawking, M.~J.~Perry and A.~Strominger,  
Phys.\ Rev.\ Lett.\ \textbf{116}, no. 23, 231301 (2016).



\bibitem{Bekenstein:1974ax}  J.~D.~Bekenstein,  
Phys.\ Rev.\ D \textbf{9}, 3292 (1974).

\bibitem{Christensen:1977jc} 
  S.~M.~Christensen and S.~A.~Fulling,
  Phys.\ Rev.\ D {\bf 15}, 2088 (1977).
  
  
  
\bibitem{Banerjee:2008cf}  R.~Banerjee and B.~R.~Majhi, 
JHEP \textbf{0806}, 095 (2008).

\bibitem{Kerner:2006vu}  R.~Kerner and R.~B.~Mann,  
Phys.\ Rev.\ D \textbf{73}, 104010 (2006).

\bibitem{Kerner:2007jk}  R.~Kerner and R.~B.~Mann,  
Phys.\ Rev.\ D \textbf{75}, 084022 (2007).


\bibitem{Parikh:1999mf}  M.~K.~Parikh and F.~Wilczek,  
Phys.\ Rev.\ Lett.\ \textbf{85}, 5042 (2000).

\bibitem{Akhmedov:2006pg}  E.~T.~Akhmedov, V.~Akhmedova and D.~Singleton,  
Phys.\ Lett.\ B \textbf{642}, 124 (2006).

\bibitem{Robinson:2005pd} 
  S.~P.~Robinson and F.~Wilczek,
  Phys.\ Rev.\ Lett.\  {\bf 95}, 011303 (2005)
  
  
  
\bibitem{Akhmedova:2010zz}  V.~E.~Akhmedova, T.~Pilling, A.~de Gill and
D.~Singleton,  
Theor.\ Math.\ Phys.\ \textbf{163}, 774 (2010).  


  
\bibitem{Kruglov1}  S.~I.~Kruglov,  
Int.\ J.\ Mod.\ Phys.\ A \textbf{29}, 1450118 (2014).

\bibitem{Kruglov2}  S.~I.~Kruglov,  
Mod.\ Phys.\ Lett.\ A \textbf{29}, no. 39, 1450203 (2014).

\bibitem{Kuang:2017sqa}  X.~M.~Kuang, J.~Saavedra and A.~\"{O}vg\"{u}n, 
  Eur.\ Phys.\ J.\ C {\bf 77}, no. 9, 613 (2017).

\bibitem{Sakalli:2017ewb} 
  I.~Sakalli and A.~Ovgun,
  Europhys.\ Lett.\  {\bf 118}, no. 6, 60006 (2017)



\bibitem{Jusufi:2017trn} 
  K.~Jusufi, I.~Sakallı and A.~\"{O}vg\"{u}n,
  Gen.\ Rel.\ Grav.\  {\bf 50}, no. 1, 10 (2018).
  
\bibitem{Sakalli:2016mnk} 
  I.~Sakalli, A.~\"{O}vg\"{u}n and K.~Jusufi,
  Astrophys.\ Space Sci.\  {\bf 361}, no. 10, 330 (2016).
  

\bibitem{Ovgun:2016roz}  A.~\"{O}vg\"{u}n,  
Advances in High Energy Physics, vol. 2017, 1573904, 9,2017.

\bibitem{Sakalli:2016cbo} 
  I.~Sakalli and A.~\"{O}vg\"{u}n,
  Eur.\ Phys.\ J.\ Plus {\bf 131}, no. 6, 184 (2016).


\bibitem{Ovgun:2015box}  A.~Ovgun and K.~Jusufi,  
Eur.\ Phys.\ J.\ Plus \textbf{131}, no. 5, 177 (2016).

\bibitem{Sakalli:2015jaa}  I.~Sakalli and A.~Ovgun,  
Gen.\ Rel.\ Grav.\ \textbf{48}, no. 1, 1 (2016).

\bibitem{Sakalli:2015taa}  I.~Sakalli and A.~Ovgun,  
Eur.\ Phys.\ J.\ Plus \textbf{130}, no. 6, 110 (2015).

\bibitem{Akhmedova1} V. Akhmedova, T. Pilling, A. de Gill, and D. Singleton, Phys. Lett. B \textbf{666}, 269 (2008).


\bibitem{aji} 
  A.~\"{O}vg\"{u}n, I.~Sakalli and J.~Saavedra,
  arXiv:1708.08331.


\bibitem{Sakalli:2016jkf}  I.~Sakalli and G.~Tokgoz,  
Annalen Phys.\ \textbf{528}, 612 (2016).

\bibitem{Sakalli:2014wja}  I.~Sakalli,  
Eur.\ Phys.\ J.\ C \textbf{75}, no. 4, 144 (2015).

\bibitem{Sakalli:2016fif} 
  I.~Sakalli,
  Phys.\ Rev.\ D {\bf 94}, no. 8, 084040 (2016)





\bibitem{Maldacena:1996ix}
  J.~M.~Maldacena and A.~Strominger,
  Phys.\ Rev.\ D {\bf 55} (1997) 861.


\bibitem{Harmark:2007jy}
T.~Harmark, J.~Natario and R.~Schiappa,
Adv.\ Theor.\ Math.\ Phys.\  {\bf 14}, 727 (2010).


\bibitem{Corda:2012tz} 
  C.~Corda,
  Int.\ J.\ Mod.\ Phys.\ D {\bf 21}, 1242023 (2012).
  
\bibitem{Corda:2012dw} 
  C.~Corda,
  Eur.\ Phys.\ J.\ C {\bf 73}, 2665 (2013).
  
  

\bibitem{Regge:1957td} T.~Regge and J.~A.~Wheeler, 
Phys.\ Rev.\ \textbf{108}, 1063 (1957).


\bibitem{Zerilli:1971wd} F.~J.~Zerilli, 
Phys.\ Rev.\ D \textbf{2}, 2141 (1970).


\bibitem{Zerilli:1970se} F.~J.~Zerilli, 
Phys.\ Rev.\ Lett.\ \textbf{24}, 737 (1970). 


\bibitem{Kokkotas:1999bd} K.~D.~Kokkotas and B.~G.~Schmidt, 
Living Rev.\ Rel.\ \textbf{2}, 2 (1999) 


\bibitem{Nollert:1999ji} H.~-P.~Nollert, 
Class.\ Quant.\ Grav.\ \textbf{16}, R159 (1999). 


\bibitem{Konoplya:2011qq} R.~A.~Konoplya and A.~Zhidenko, 
Rev.\ Mod.\ Phys.\ \textbf{83}, 793 (2011).




\bibitem{Maldacena:1997re} J.~M.~Maldacena, 
Adv.\ Theor.\ Math.\ Phys.\ \textbf{2}, 231 (1998). 


\bibitem{Horowitz:1999jd} G.~T.~Horowitz and V.~E.~Hubeny, 
Phys.\ Rev.\ D \textbf{62}, 024027 (2000).


\bibitem{Abbott:2016blz}
  B.~P.~Abbott {\it et al.} [LIGO Scientific and Virgo Collaborations],
  Phys.\ Rev.\ Lett.\  {\bf 116}, no. 6, 061102 (2016)

\bibitem{TheLIGOScientific:2016src}
  B.~P.~Abbott {\it et al.} [LIGO Scientific and Virgo Collaborations],
  Phys.\ Rev.\ Lett.\  {\bf 116}, no. 22, 221101 (2016)

\bibitem{Konoplya:2016pmh}
  R.~Konoplya and A.~Zhidenko,
  Phys.\ Lett.\ B {\bf 756}, 350 (2016)




\bibitem{Deser:1981wh}  S.~Deser, R.~Jackiw and S.~Templeton,  
Annals Phys.\ \textbf{140} (1982) 372  [Erratum-ibid.\ \textbf{185} (1988)
406]  [Annals Phys.\ \textbf{185} (1988) 406]  [Annals Phys.\ \textbf{281}
(2000) 409].  


\bibitem{Deser:1982vy}  S.~Deser, R.~Jackiw and S.~Templeton,  
Phys.\ Rev.\ Lett.\ \textbf{48} (1982) 975.  



\bibitem{Garbarz:2008qn} 
  A.~Garbarz, G.~Giribet and Y.~Vasquez,
  Phys.\ Rev.\ D {\bf 79}, 044036 (2009).



\bibitem{Nakasone:2009bn}  M.~Nakasone and I.~Oda,  
Prog.\ Theor.\ Phys.\ \textbf{121} (2009) 1389.



\bibitem{Bergshoeff:2009aq}  E.~A.~Bergshoeff, O.~Hohm and P.~K.~Townsend,  
Phys.\ Rev.\ D \textbf{79} (2009) 124042.  


\bibitem{Oda:2009ys}  I.~Oda,  
JHEP \textbf{0905} (2009) 064.  


 \bibitem{Ohta:2011rv} 
  N.~Ohta,
  Class.\ Quant.\ Grav.\  {\bf 29}, 015002 (2012).
  
 \bibitem{Muneyuki:2012ur} 
  K.~Muneyuki and N.~Ohta,
  Phys.\ Rev.\ D {\bf 85}, 101501 (2012).
  
\bibitem{Vasquez:2009mk} 
  Y.~Vasquez,
  JHEP {\bf 1108}, 089 (2011).  



\bibitem{Banados:2005da} 
  M.~Banados, G.~Barnich, G.~Compere and A.~Gomberoff,
  Phys.\ Rev.\ D {\bf 73}, 044006 (2006).
 
  
  
  
\bibitem{Biswas:2003ku} 
  R.~Biswas, E.~Keski-Vakkuri, R.~G.~Leigh, S.~Nowling and E.~Sharpe,
  JHEP {\bf 0401}, 064 (2004)
  
\bibitem{Brecher:2003rv} 
  D.~Brecher, P.~A.~DeBoer, D.~C.~Page and M.~Rozali,
  JHEP {\bf 0310}, 031 (2003)
  
\bibitem{Brace:2003st} 
  D.~Brace, C.~A.~R.~Herdeiro and S.~Hirano,
  Phys.\ Rev.\ D {\bf 69}, 066010 (2004)
  
\bibitem{Takayanagi:2003ps} 
  H.~Takayanagi,
  JHEP {\bf 0312}, 011 (2003)
  
  



\bibitem{Cardoso:2001hn} 
  V.~Cardoso and J.~P.~S.~Lemos,
  Phys.\ Rev.\ D {\bf 63}, 124015 (2001)
 
\bibitem{Birmingham:2001pj} 
  D.~Birmingham, I.~Sachs and S.~N.~Solodukhin,
  Phys.\ Rev.\ Lett.\  {\bf 88}, 151301 (2002)
  
\bibitem{Konoplya:2004ik} 
  R.~A.~Konoplya,
  Phys.\ Rev.\ D {\bf 70}, 047503 (2004)
  
\bibitem{Kwon:2011ey} 
  Y.~Kwon, S.~Nam, J.~D.~Park and S.~H.~Yi,
  Class.\ Quant.\ Grav.\  {\bf 28}, 145006 (2011)
  
\bibitem{CuadrosMelgar:2011up} 
  B.~Cuadros-Melgar, J.~de Oliveira and C.~E.~Pellicer,
  Phys.\ Rev.\ D {\bf 85}, 024014 (2012)
  
\bibitem{Becar:2013qba} 
  R.~Becar, P.~A.~Gonzalez and Y.~Vasquez,
  Phys.\ Rev.\ D {\bf 89}, no. 2, 023001 (2014)
  
  
  
\bibitem{Gonzalez:2014voa} 
  P.~A.~Gonzalez and Y.~Vasquez,
  Eur.\ Phys.\ J.\ C {\bf 74}, no. 7, 2969 (2014)
  
\bibitem{Catalan:2014una} 
  M.~Catalan and Y.~Vasquez,
  Phys.\ Rev.\ D {\bf 90}, no. 10, 104002 (2014).
  
\bibitem{Gonzalez:2017ptj} 
  P.~A.~Gonzalez, Y.~Vasquez and R.~N.~Villalobos,
  Eur.\ Phys.\ J.\ C {\bf 77}, no. 9, 579 (2017).
  
  
\bibitem{Godel:1949ga} 
  K.~Godel,
  Rev.\ Mod.\ Phys.\  {\bf 21}, 447 (1949).



  \bibitem{Moussa:2008sj} 
  K.~A.~Moussa, G.~Clement, H.~Guennoune and C.~Leygnac,
  Phys.\ Rev.\ D {\bf 78}, 064065 (2008).


\bibitem{Gauntlett:2002nw} 
  J.~P.~Gauntlett, J.~B.~Gutowski, C.~M.~Hull, S.~Pakis and H.~S.~Reall,
  Class.\ Quant.\ Grav.\  {\bf 20}, 4587 (2003)

  
\bibitem{Herdeiro:2002ft} 
  C.~A.~R.~Herdeiro,
  Nucl.\ Phys.\ B {\bf 665}, 189 (2003). 
  
  
\bibitem{Gimon:2003ms} 
  E.~G.~Gimon and A.~Hashimoto,
  Phys.\ Rev.\ Lett.\  {\bf 91}, 021601 (2003).
  
  
\bibitem{Brecher:2003wq} 
  D.~Brecher, U.~H.~Danielsson, J.~P.~Gregory and M.~E.~Olsson,
  JHEP {\bf 0311}, 033 (2003).
  
  \bibitem{Herdeiro:2003un} 
  C.~A.~R.~Herdeiro,
  Class.\ Quant.\ Grav.\  {\bf 20}, 4891 (2003).

\bibitem{Behrndt:2004pn} 
  K.~Behrndt and D.~Klemm,
  Class.\ Quant.\ Grav.\  {\bf 21}, 4107 (2004).

\bibitem{Boyda:2002ba} 
  E.~K.~Boyda, S.~Ganguli, P.~Horava and U.~Varadarajan,
  Phys.\ Rev.\ D {\bf 67}, 106003 (2003)
  
\bibitem{Harmark:2003ud} 
  T.~Harmark and T.~Takayanagi,
  Nucl.\ Phys.\ B {\bf 662}, 3 (2003)  
  
 \bibitem{Konoplya:2005sy} 
  R.~A.~Konoplya and E.~Abdalla,
  Phys.\ Rev.\ D {\bf 71}, 084015 (2005)
 
\bibitem{Konoplya:2011ig} 
  R.~A.~Konoplya and A.~Zhidenko,
  Phys.\ Rev.\ D {\bf 84}, 064028 (2011).
    
  
\bibitem{Li:2012ee}
  R.~Li,
  Phys.\ Rev.\ D {\bf 85} (2012) 065020
  
  

 \bibitem{Barnich:2005kq} 
  G.~Barnich and G.~Compere,
  Phys.\ Rev.\ Lett.\  {\bf 95}, 031302 (2005).
 

\bibitem{Konoplya:2011it} 
  R.~A.~Konoplya and A.~Zhidenko,
  Phys.\ Rev.\ D {\bf 84}, 104022 (2011).
  
  
\bibitem{Konoplya:2011ag} 
  R.~A.~Konoplya,
  Phys.\ Lett.\ B {\bf 706}, 451 (2012).
  
  
  
\bibitem{Li:2012qc} 
  R.~Li,
  Int.\ J.\ Mod.\ Phys.\ D {\bf 21}, 1250014 (2012)
  



  



\bibitem{Akhmedov:2006un}  E.~T.~Akhmedov, V.~Akhmedova, T.~Pilling and
D.~Singleton,  
Int.\ J.\ Mod.\ Phys.\ A \textbf{22}, 1705 (2007).





\bibitem{Abramowitz} M. Abramowitz and A. Stegun, Handbook of
Mathematical functions, (Dover publications, New York, 1970).



\bibitem{Pavan:2009wt} 
  A.~B.~Pavan, E.~Abdalla and C.~Molina,
  Phys.\ Rev.\ D {\bf 81}, 044003 (2010)



\bibitem{ch1} 
  S.~Chandrasekhar,
  Proc.\ Roy.\ Soc.\ Lond.\ A {\bf 349}, 571 (1976).
  
  \bibitem{ch2} 
  S.~Chandrasekhar,
  OXFORD, UK: CLARENDON (1985) 646 P.


\bibitem{Gubser:1997qr} 
  S.~S.~Gubser,
  Phys.\ Rev.\ D {\bf 56}, 4984 (1997).






\bibitem{Benone:2015bst}
  C.~L.~Benone and L.~C.~B.~Crispino,
  Phys.\ Rev.\ D {\bf 93}, no. 2, 024028 (2016).

\bibitem{Gonzalez:2017shu} 
  P.~A.~Gonzalez, E.~Papantonopoulos, J.~Saavedra and Y.~Vasquez,
  Phys.\ Rev.\ D {\bf 95}, no. 6, 064046 (2017).
  


\bibitem{Gonzalez:2010vv} 
  P.~Gonzalez, E.~Papantonopoulos and J.~Saavedra,
  JHEP {\bf 1008}, 050 (2010).
  
  
\bibitem{Gonzalez:2010ht} 
  C.~Campuzano, P.~Gonzalez, E.~Rojas and J.~Saavedra,
  JHEP {\bf 1006}, 103 (2010).
  
  
\bibitem{Gonzalez:2011du} 
  P.~A.~Gonzalez and J.~Saavedra,
  Int.\ J.\ Mod.\ Phys.\ A {\bf 26}, 3997 (2011).
  
  
\bibitem{Catalan:2014ama} 
  M.~Catalan, E.~Cisternas, P.~A.~Gonzalez and Y.~Vasquez,
  Astrophys.\ Space Sci.\  {\bf 361}, no. 6, 189 (2016).
  
\bibitem{Becar:2014aka} 
  R.~Becar, P.~A.~Gonzalez and Y.~Vasquez,
  Eur.\ Phys.\ J.\ C {\bf 74}, no. 8, 3028 (2014).
  
  
  
\bibitem{Becar:2014saa} 
  R.~Becar, P.~A.~Gonzalez, J.~Saavedra and Y.~Vasquez,
  Eur.\ Phys.\ J.\ C {\bf 75}, no. 2, 57 (2015).




\end{thebibliography}
\end{document}